\documentclass[a4paper,aps,prb,twocolumn,amsmath,superscriptaddress]{revtex4-2}
\usepackage{graphicx}
\usepackage{color}
\usepackage{epstopdf}

\usepackage{xcolor}

\begin{document}

\title{Strain-driven spin mixing and dark-exciton recombination in a neutral Ni$^{2+}$-doped quantum dot}

\author{K. E.~Polczynska}
\affiliation{Univ. Grenoble Alpes, CNRS, Grenoble INP, Institut N\'{e}el, 38000 Grenoble, France}
\affiliation{Univ. of Warsaw, Faculty of Physics, ul. Pasteura 5, 02-093 Warsaw, Poland}
\author{S.~Karouaz}
\affiliation{Univ. Grenoble Alpes, CNRS, Grenoble INP, Institut N\'{e}el, 38000 Grenoble, France}
\author{W.~Pacuski}
\affiliation{Univ. of Warsaw, Faculty of Physics, ul. Pasteura 5, 02-093 Warsaw, Poland}
\author{L.~Besombes}\email{lucien.besombes@neel.cnrs.fr}
\affiliation{Univ. Grenoble Alpes, CNRS, Grenoble INP, Institut N\'{e}el, 38000 Grenoble, France}

\date{\today}

\begin{abstract}

We investigate the optical properties of neutral excitons in CdTe/ZnTe quantum dots containing a single Ni$^{2+}$ ion. We show that the photoluminescence spectra provide a direct spectroscopic signature of strain-induced mixing of the Ni$^{2+}$ spin states. A misalignment between the principal axis of the local strain tensor and the quantum dot growth direction reorients the spin quantization axis of the magnetic ion, reducing the hole–Ni$^{2+}$ exchange interaction at low magnetic field and giving rise to photoluminescence replicas around the partially linearly polarized bright-exciton transitions. A longitudinal magnetic field restores the circularly polarized optical selection rules, allowing the three spin projections $S_z = 0, \pm 1$ of the Ni$^{2+}$ ion to be spectrally resolved. Dark-exciton emission appears on the low-energy side of the spectra and is dominated at low field by transitions involving spin flips of the magnetic ion. An effective spin Hamiltonian including strain orientation and valence-band mixing reproduces the magnetic-field evolution of both bright- and dark-exciton spectra. These results highlight the key role of the local strain environment in determining the spin–exciton coupling of transition-metal dopants in semiconductor quantum dots.

\end{abstract}

\maketitle

\section{Introduction}

Individual optically active magnetic defects in solid-state materials constitute quantum systems that can operate as qubits with a spin–photon interface. The properties of a given defect strongly depend on its host environment, leading to specific optical selection rules and control schemes. Among these systems, transition-metal ions incorporated in semiconductors are promising for quantum applications \cite{Awschalom2020,Bacher2016,Krebs2025}. Their spin and optical properties originate from the partially filled 3$d$ shell, which typically interacts with the free carriers of the host material \cite{Furdyna1988}.

A broad range of magnetic elements can be embedded in conventional semiconductors, offering different combinations of electronic spin, nuclear spin, and orbital momentum. Transition-metal ions with nonzero orbital momentum are particularly sensitive to the crystal field and to its modification by local strain \cite{Kudelski2007,Lafuente2016}. While some of these ions exhibit intrinsic optical transitions, semiconductor quantum dots (QDs) provide an alternative approach in which the confined carriers probe and control the magnetic dopant via exchange interaction. Neutral CdTe/ZnTe QDs with narrow excitonic lines in the visible range (around 600 nm) have thus enabled optical investigations of single Mn, Co, Cr, and V spins \cite{Besombes2004,Kobak2014,Lafuente2016,Polczynska2025}.

Here we focus on nickel incorporated in a II–VI semiconductor as a Ni$^{2+}$ ion (3d$^8$, $S=1$, $L=3$). Owing to its finite orbital momentum and spin–orbit coupling, Ni$^{2+}$ exhibits a pronounced spin–strain coupling. This property makes the $S=1$ system particularly attractive for hybrid spin–mechanical platforms, where the nearly degenerate $S_z=\pm1$ states could be coherently coupled by dynamical in-plane strain \cite{Besombes2019,Tiwari2020JAP,Barfuss2015}. The influence of local strain on the emission of positively charged Ni$^{2+}$-doped QDs has recently been reported \cite{PRB2025}. For spin–mechanical applications, however, a neutral QD is required, ensuring that the Ni$^{2+}$ spin is decoupled from carriers in the absence of optical excitation.  

In this work, we investigate the optical and spin properties of neutral Ni$^{2+}$-doped CdTe/ZnTe QDs. This neutral configuration allows the intrinsic strain-induced mixing of the Ni$^{2+}$ spin states and its influence on both bright and dark excitons to be investigated. We show that the orientation of the local strain tensor can strongly modify the effective spin quantization axis of the Ni$^{2+}$ ion, leading to characteristic signatures in the exciton photoluminescence (PL) spectra. A misalignment between the principal strain axis and the QD growth direction reduces the hole–Ni$^{2+}$ exchange interaction at zero or weak magnetic field and induces PL replicas on both sides of the bright-exciton transitions. A longitudinal magnetic field progressively restores the hole–Ni$^{2+}$ exchange interaction and the circularly polarized optical selection rules, enabling spectral resolution of the three spin projections $S_z = 0, \pm1$ in the bright-exciton emission at fields of a few Tesla. Dark excitons are observed on the low-energy side of the spectra; at low magnetic field their emission is dominated by spin-flip replicas involving changes in the Ni$^{2+}$ spin levels.

The remainder of this paper is organized as follows. In Sec.~II, we describe the sample and experimental methods and present the zero-field emission of a neutral Ni$^{2+}$-doped QD, followed by its longitudinal magnetic-field dependence for both bright and dark excitons. In Sec.~III, we introduce an effective spin Hamiltonian showing that strain misorientation at the magnetic-ion site accounts for the main spectral features. In Sec.~IV, we present modeling results that, in particular, clarify the mechanisms enabling dark-exciton recombination. Conclusions are given in Sec.~V.

\section{Optical probing of neutral Ni$^{2+}$-doped quantum dots}

\subsection{Sample and experiments}

The Ni$^{2+}$-doped, self-assembled CdTe/ZnTe QDs \cite{Wojnar2011,Kobak2013} investigated in this work were grown by molecular beam epitaxy, following procedures similar to those used for previously developed CdTe/ZnTe single-spin systems (see, e.g., Refs.~\cite{Besombes2004,Kobak2014,Lafuente2016,Polczynska2025}). A 100 nm ZnTe buffer layer and a 3 µm CdTe layer were first deposited on a GaAs (100) substrate with a 2° miscut. A 660 nm ZnTe bottom barrier was then grown, followed by the deposition of 1.5 nm CdTe, leading to the formation of Stranski–Krastanov QDs. The substrate temperature was initially set to 370°C and reduced to 350°C for the QD growth step. A low Ni flux was supplied throughout the QD deposition \cite{PRB2025}

Single QDs are investigated by optical micro-spectroscopy at liquid helium temperature (T = 4.2 K). The sample is mounted on x–y–z piezoelectric stages (Attocube ANPx(z)101) inside a vacuum tube under low-pressure He exchange gas, and immersed in the variable-temperature insert of a liquid-He cryostat. The temperature is monitored by a sensor integrated into the sample holder. The cryostat includes a vector superconducting magnet, allowing magnetic fields up to 9 T along the QD growth axis.

PL is excited using a tunable dye laser (Coherent CR599 with Rhodamine 6G) and collected through a high-numerical-aperture microscope objective (NA = 0.85). For spectral analysis, the emission is dispersed by a 2 m double-grating spectrometer (Jobin Yvon U1000, 1800 gr/mm gratings), providing a resolution of ~25 µeV in the relevant spectral range, and detected with a cooled Si CCD camera (Andor Newton).

For PL excitation (PLE) measurements, the laser power is stabilized with an electro-optic variable attenuator (Cambridge Research \& Instrumentation LPC) during wavelength tuning. Linear polarization is analyzed using a half-wave plate (B. Halle, 460–680 nm) followed by a Glan–Taylor polarizer (Thorlabs GT10-A). Circular polarization is measured by inserting a quarter-wave plate (B. Halle, 460–680 nm) at 45° relative to the detection polarization axis.

\subsection{Photo-luminescence of X-Ni$^{2+}$}

\begin{figure}[hbt]
\centering
\includegraphics[width=1\linewidth]{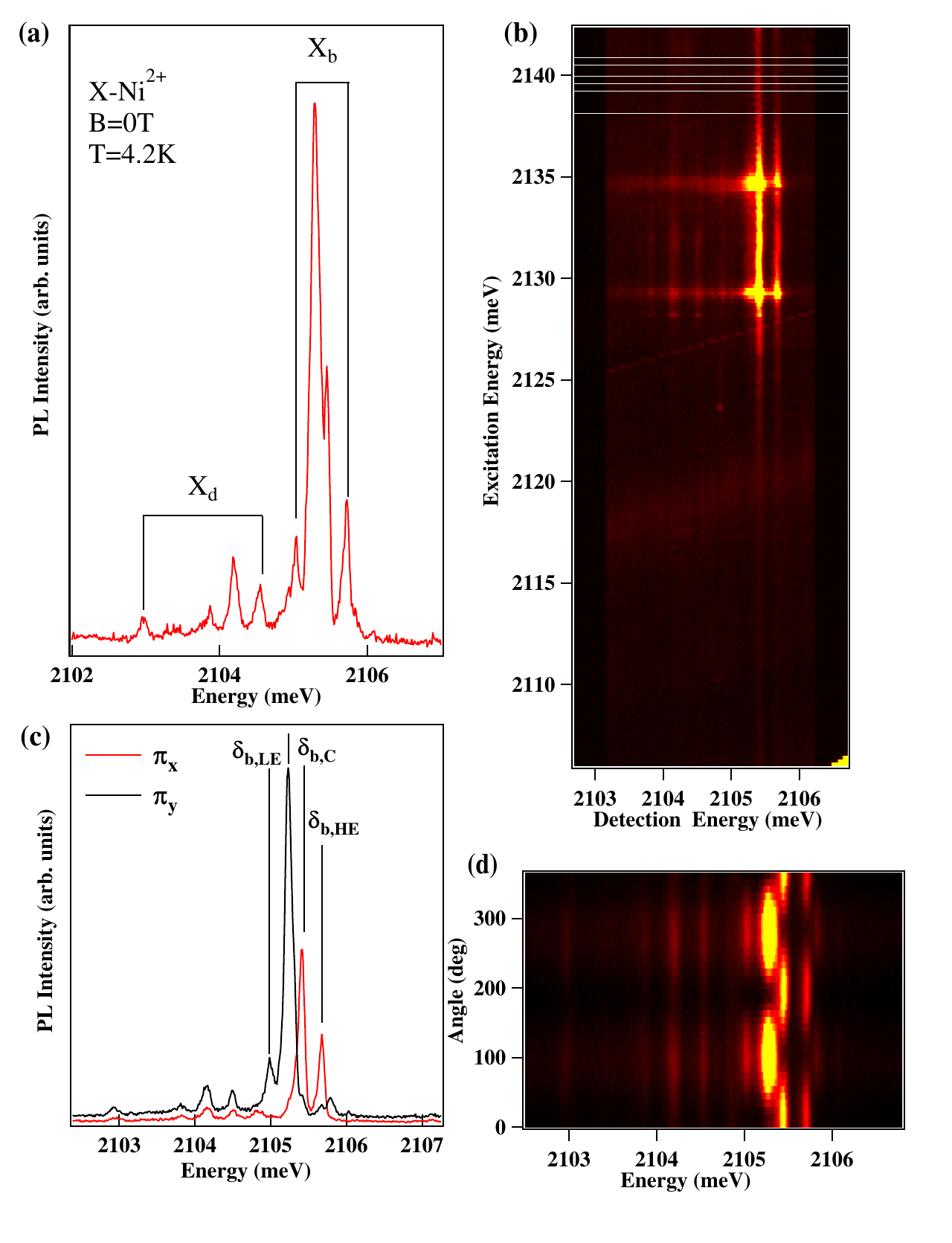}
\caption{(a) PL spectra of the exciton in a Ni$^{2+}$-doped QD at B$_z$=0T for an excitation on an excited state at E = 2129.3 meV. (b) PL excitation spectra of the neutral exciton. (c) Intensity map of the linearly polarized PL of the exciton at B$_z$=0~T. (d) Linearly polarized PL spectra recorded along two orthogonal directions.}
\label{Fig:XB0}
\end{figure}

The QDs in the studied Ni-doped sample are often positively charged \cite{PRB2025}; however, emission from the neutral exciton (X) can be observed in some cases for optical excitation energies below the band gap of the ZnTe barriers. The PL and PLE spectra of a neutral exciton in a Ni$^{2+}$-doped QD are shown in Fig.~\ref{Fig:XB0}. The X–Ni$^{2+}$ complex can be investigated under specific excitation conditions involving excited states of the dot (Fig.~\ref{Fig:XB0}(b)). As shown in Fig.~\ref{Fig:XB0}(a), the PL spectrum of X–Ni$^{2+}$ is dominated by two broad, partially overlapping lines. The linear polarization properties of X–Ni$^{2+}$ at zero magnetic field are presented in Fig.~\ref{Fig:XB0}(c) and \ref{Fig:XB0}(d). The two main PL lines are linearly polarized along orthogonal directions. Weaker satellite lines are observed on both sides of the main optical transitions; these lines are also linearly polarized along the same directions as the corresponding main transitions.

In neutral, non-magnetic QDs, the electron–hole exchange interaction splits the exciton into two bright states (with anti-parallel electron and hole spins) and two dark states (with parallel electron and hole spins). In the presence of in-plane anisotropy (due to shape and/or strain), the bright states are mixed, and their emission becomes linearly polarized along orthogonal directions \cite{Bayer2002}. In a magnetic QD, the four exciton states interact with the Ni$^{2+}$ ion spin and the two main broad, linearly polarized lines in Fig.~\ref{Fig:XB0}(c) likely correspond to the bright exciton states.

Additional weaker lines, labeled X$_{d}$ in Fig.~\ref{Fig:XB0}(a), appear on the low-energy side of the PL spectrum. These weak-intensity lines exhibit partial linear polarization aligned with the low-energy, high-intensity bright exciton line (Fig.~\ref{Fig:XB0}(d)). As will be shown later through magneto-optical measurements, they can be attributed to dark exciton recombination.

\subsection{Magneto-optical properties of X-Ni$^{2+}$}

\begin{figure}[hbt]
\centering
\includegraphics[width=1.05\linewidth]{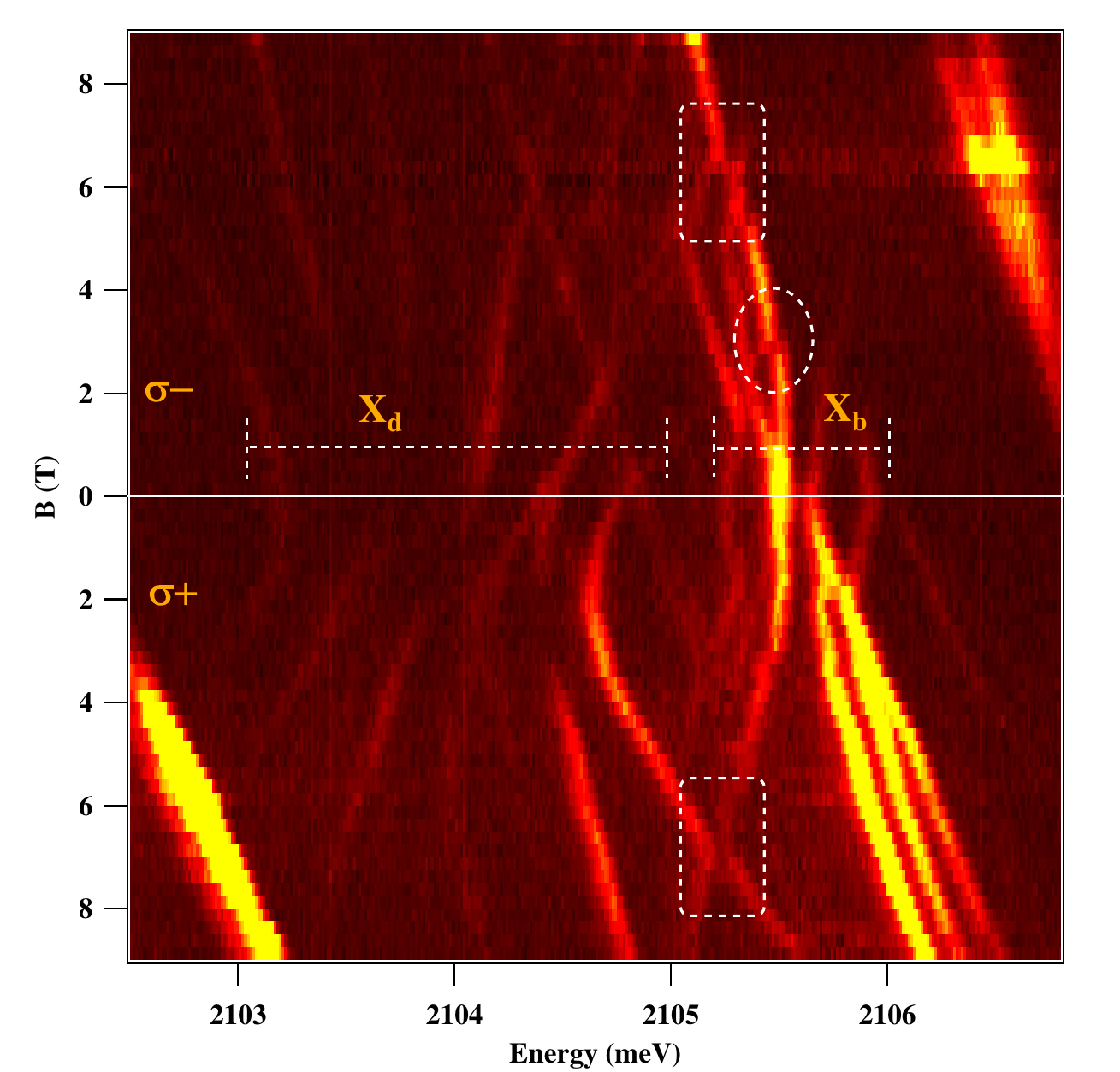}
\caption{Circularly polarized PL intensity map of the longitudinal magnetic field (B$_z$) dependence of the bright (X$_b$) and dark (X$_d$) excitons in Ni$^{2+}$-doped CdTe/ZnTe QD. The fan-like structure of the dark exciton transitions reflects recombination processes involving changes in the Ni$^{2+}$ spin levels.}
\label{Fig:MapBX}
\end{figure}

The dependence of the X–Ni$^{2+}$ emission on a longitudinal magnetic field ($B_z$) is presented in Fig.~\ref{Fig:MapBX}. Under $B_z$, the two main optical transitions split, in each circular polarization, into a three-line structure. This is particularly evident in $\sigma^+$ polarization, where the lines shift toward higher energy and remain unaffected by overlap with lower-energy optical transitions. In $\sigma^-$ polarization, the emission pattern is more complex: anti-crossings with lower-energy lines appear as $B_z$ increases. The six high-intensity lines observed under $B_z$ — three in each circular polarization — correspond to the two bright exciton states interacting with the three spin projections of the Ni$^{2+}$ ion, $S_z = 0, \pm 1$ (with z the QD growth axis).

The lower-energy, weak-intensity lines labeled X$_d$ evolve into a wide fan under $B_z$, indicating that some of these transitions are characterized by an effective $g$-factor larger than that of a standard exciton. These optical transitions likely occur together with a change in the spin state of the magnetic atom, so that their magnetic-field dependence involves the $g$-factor of the Ni$^{2+}$ ion. This group of lines, centered approximately at $\delta_{bd} \approx 1.65$~meV below the center of the bright exciton transitions, originates from dark excitons exchange-coupled to the Ni$^{2+}$ spin.

\begin{table}[!hbt]
\center
\begin{tabular}{c|ccccccccccccc}
\hline \hline
X    & $\delta_{bd}$ & $\delta_d$(0T) & $\delta_d$(min) & $\delta_{b,LE}$ & $\delta_{b,C}$ & $\delta_{b,HE}$  &$\delta_b$(0T) & $\delta_{\pm 1}$(9T)\\
\hline
QD1  & 1.65           &1.55           & 1.47              & 0.23             & 0.14            & 0.27              & 0.64          & 0.34 \\
\hline \hline       
\end{tabular}
\caption{Energy splittings (in meV) extracted from the PL spectra shown in Fig.~\ref{Fig:XB0} and Fig.~\ref{Fig:MapBX}. Here, $\delta_{bd}$ denotes the energy separation between the centers of the bright- and dark-exciton manifolds; $\delta_d$ and $\delta_b$ are the total splittings of the dark and bright excitons, respectively; $\delta_{b,i}$ represents the splittings between the bright-exciton lines at $B_z = 0$; and $\delta_{\pm1}$ is the splitting between the $S_z = \pm1$ bright-exciton lines at $B_z = 9$T in $\sigma^+$ polarization.}
\label{TableSplit}
\end{table}

\section{Bright excitons in a Ni$^{2+}$-doped quantum dot}

The emission spectra observed in a neutral Ni$^{2+}$-doped QD at zero or weak magnetic field differ significantly from the simple three-line structure expected for a spin $S = 1$ interacting with an exciton. An effective spin model can be employed to identify the key parameters governing these complex emission spectra.

\subsection{Model of X-Ni$^{2+}$}

The model recently introduced in Ref.~\cite{PRB2025} to describe a positively charged exciton coupled to a Ni$^{2+}$ spin can be extended to calculate the X–Ni$^{2+}$ spectra. In the optical recombination of X–Ni$^{2+}$, the ground state corresponds to an empty dot whose energy levels are governed solely by the zero-field splitting of the Ni$^{2+}$ ion and its Zeeman interaction, $\mathcal{H}_{Ni}=\mathcal{H}_{ZFS}+g_{Ni}\mu_B\vec{S}.\vec{B}$ with

\begin{eqnarray}
\mathcal{H}_{ZFS}=\frac{2}{3}D_{0}\left[S_{z^{\prime \prime}}^2-\frac{1}{2}\left(S_{x^{\prime \prime}}^2+S_{y^{\prime \prime}}^2\right)\right]+\frac{E}{2}\left(S_{x^{\prime \prime}}^2-S_{y^{\prime \prime}}^2\right)
\label{ZFS} 
\end{eqnarray}

\noindent where $D_0$ and $E$ are strain-induced fine-structure parameters \cite{PRB2025,Krebs2009} arising from modifications of the crystal field and spin–orbit coupling. For a general description accounting for a possible misalignment of the local strain tensor with respect to the QD axes, $(x^{\prime\prime}, y^{\prime\prime}, z^{\prime\prime})$ defines the local strain frame at the position of the magnetic atom. This frame is specified by three angles $\vartheta_s$, $\varphi_s$, and $\psi_s$, and is obtained by rotating the crystal frame $(x,y,z)$ by an angle $\vartheta_s$ around the unitary vector $\vec{u_s}(-sin(\varphi_s)/a,cos(\varphi_s)/a,sin(\psi_s)/a)$ with $a=\sqrt{\sin^2(\varphi_s)+\cos^2(\varphi_s)+\sin^2(\psi_s)}$ \cite{PRB2025,Krebs2009}. For an arbitrary orientation of the local strain tensor, $\mathcal{H}_{ZFS}$ mixes the three $S_z$ spin projections, resulting in three non-degenerate energy levels in the ground state of an empty Ni$^{2+}$-doped dot. Such a misalignment of the strain tensor is expected for magnetic ions incorporated at non-central positions in self-assembled QDs, where local lattice relaxation and alloy disorder can significantly modify the crystal field experienced by the dopant.

In the excited state, the Hamiltonian of the X–Ni$^{2+}$ complex reads

\begin{eqnarray}
\mathcal{H}_{X,Ni}=E_g+\mathcal{H}_{VB}+\mathcal{H}_{e-B}+\mathcal{H}_{h-B}\\ \nonumber
+\mathcal{H}_{e-Ni}+\mathcal{H}_{h-Ni}+\mathcal{H}_{eh}+\mathcal{H}_{Ni}
\label{Hamilton}
\end{eqnarray}

\noindent where $E_g$ is the electron energy, $\mathcal{H}_{VB}$ describes the heavy-hole and light-hole ground states including possible valence-band mixing (see Appendix A), and $\mathcal{H}_{i-B}$ denotes the Zeeman interaction of the carriers. $\mathcal{H}_{i-Ni}$ accounts for the carrier–Ni$^{2+}$ exchange interaction. For the electron, $\mathcal{H}_{e-Ni}=I_{eNi}.\vec{\sigma}\cdot\vec{S}$ describes the ferromagnetic electron–Ni$^{2+}$ coupling. In magnetic QDs, it has been shown that the coupling with carriers is usually dominated by the kinetic exchange interaction with the hole \cite{Kacman1992,Kacman2001,Rossier2006}. This coupling is typically anti-ferromagnetic, as was found in charged Ni$^{2+}$-doped QDs. It is also much larger than the coupling with the electron, as reported in  \cite{PRB2025}.

The hole–Ni$^{2+}$ interaction $\mathcal{H}_{h-Ni}$ includes the usual isotropic Heisenberg term, corresponding to the spherical symmetry approximation, $\mathcal{H}_{h-Ni}=I_{hNi}\vec{J_h}\cdot\vec{S}$. Here, $I_{hNi}$ is the exchange integral, which depends on the amplitude of the hole envelope function at the position of the magnetic atom. Neglecting possible valence-band mixing, the hole ground state in a CdTe/ZnTe QD is a heavy hole with its spin aligned along the growth axis $z$. The exchange interaction with the hole therefore acts on the magnetic ion as an effective magnetic field oriented along $z$, $B_{ex,hNi}=I_{hNi}J_z/g_{Ni}\mu_B$ where $J_z$ is the hole spin projection \cite{Leger2005}.

It has been demonstrated both theoretically and experimentally \cite{Bhatta2003,Bhatta2007,Krebs2009,PRB2025} that when the magnetic atom is not located at the center of the dot, the hole–Ni$^{2+}$ exchange interaction can be affected by mixing between the heavy-hole ground state and higher-energy hole states. This mixing is induced by the off-diagonal terms of the Kohn–Luttinger Hamiltonian and exists even in a spherical confinement potential \cite{Bhatta2003}.

Such subband mixing introduces, in particular, an additional heavy-hole spin-flip processes that conserve the Ni$^{2+}$ spin, $\langle\pm3/2\vert\mathcal{H}_{h-Ni}\vert\mp3/2\rangle=\eta I_{hNi}S_z$ \cite{PRB2025}. The strength of this mechanism depends on the exchange integral $I_{hNi}$ and on a position-dependent parameter $\eta$. This additional spin-flip term has no direct counterpart in an anisotropic QD (shape and/or strain anisotropy) described by $\mathcal{H}_{VB}$. In anisotropic dots, in-plane anisotropy enables hole–Ni$^{2+}$ flip-flop processes, while vertical distortion allows spin flips of the magnetic atom that conserve the heavy-hole spin (Appendix A).

To describe the neutral exciton, the electron–hole exchange interaction, $\mathcal{H}_{eh}$, must also be included in the Hamiltonian. It consists of a short-range and a long-range contribution. We consider here the simplest case of QDs with $C_{2v}$ symmetry (e.g., flat ellipsoidal lenses), where both short-range and long-range terms contribute to the bright–dark exciton splitting $\delta_0$, while the long-range term additionally induces a fine-structure splitting $\delta_1$ of the bright excitons \cite{Lafuente2016}. The electron–hole exchange interaction thus leads to four exciton states which, when coupled to the three spin projections of the magnetic atom, can give rise to up to twelve energy levels.

\subsection{Longitudinal magnetic field dependence of X-Ni$^{2+}$}

Optical recombination of an exciton preserves the Ni$^{2+}$ spin, and up to six emission lines are therefore expected for the bright X–Ni$^{2+}$ complex. However, for a misaligned local strain tensor, the Ni$^{2+}$ spin states are mixed by $\mathcal{H}_{Ni}$, and the eigenstates become linear combinations of $S_z = 0$ and $S_z = \pm 1$. When the angle between the principal axis of the strain tensor and the QD growth axis is large, strong spin mixing occurs: the expectation value of $S_z$ in each eigenstate can approach zero, and the exchange interaction with the exciton — mainly governed by the coupling to the heavy-hole spin aligned along $z$ — is reduced.

In the presence of in-plane anisotropy (shape and/or strain), the bright excitons are mixed and split by the electron–hole exchange interaction, leading to the exciton fine-structure splitting. When this fine-structure splitting exceeds the exciton–Ni$^{2+}$ exchange interaction, two linearly polarized PL lines, slightly broadened by the residual coupling to the magnetic atom, are expected. The two broad, linearly polarized lines observed at zero magnetic field for the bright exciton in Fig.~\ref{Fig:XB0} result from the combined effect of a weak exciton–Ni$^{2+}$ exchange interaction and the electron–hole exchange interaction.

However, in the excited state, the effective exchange field of the hole modifies the strain-induced mixing of the $S_z$ spin projections. The Ni$^{2+}$ eigenstates therefore slightly differ in the ground and excited states. In the ground state, the spin mixing is solely governed by $\mathcal{H}_{Ni}$, whereas in the excited state, the exchange interaction with the heavy-hole spin tends to align the Ni$^{2+}$ spin along the $z$ axis, partially suppressing the strain-induced mixing. This difference in the composition of the Ni$^{2+}$ eigenstates enables optical recombination of X$_b$ accompanied by a change in the Ni$^{2+}$ spin level. The amplitude of these transitions is determined by the modification of the $S_z$ components in the mixed levels. 

These recombination processes give rise to the weak-intensity lines observed on both sides of the main bright-exciton emission: at +0.27 meV on the high-energy side and -0.24 meV on the low-energy side (see Table~\ref{TableSplit} for the detailed energy splittings). Their spectral positions are governed by the strain-induced fine-structure parameters of $\mathcal{H}_{Ni}$, namely $D_0$, $E$, and the rotation angle of the local strain frame with respect to the QD growth axis, primarily $\vartheta_s$.

\begin{figure}[hbt]
\centering
\includegraphics[width=1\linewidth]{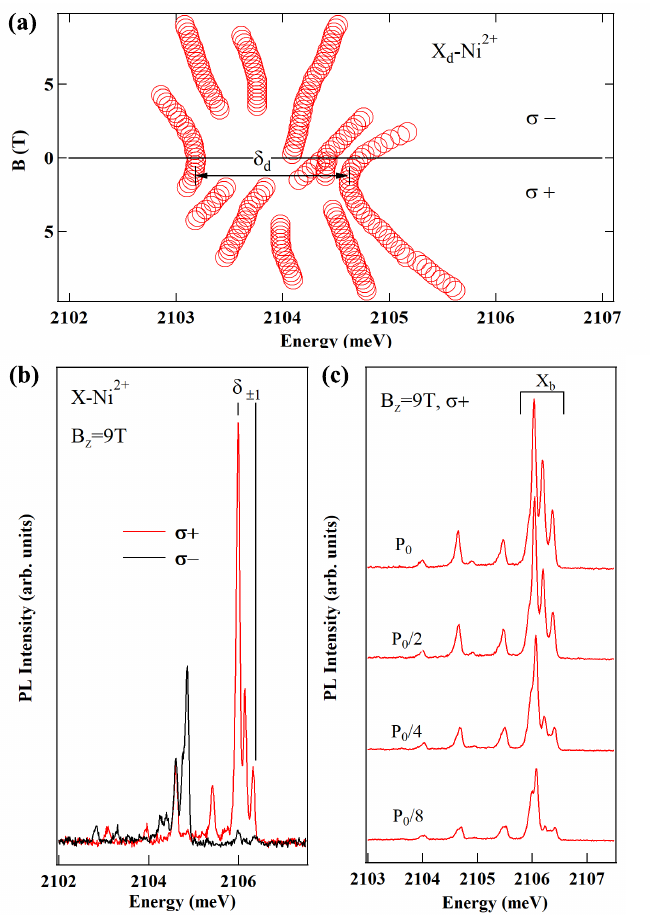}
\caption{(a) Detailed magnetic-field dependence of the dark-exciton PL transition energies in a Ni$^{2+}$-doped QD. (b) Circularly polarized PL spectra of the neutral exciton in a Ni$^{2+}$-doped QD at $B_z = 9$ T. (c) Excitation-power dependence of the exciton PL intensity in $\sigma^+$ polarization at $B_z = 9$ T}
\label{Fig:PW}
\end{figure}

When a longitudinal magnetic field $B_z$ is applied, the Zeeman energy of the Ni$^{2+}$ ion progressively increases and can eventually overcome the strain-induced mixing of the spin states. At sufficiently large $B_z$, the Ni$^{2+}$ spin aligns parallel to the magnetic field, i.e., along the QD growth axis. The quantization axis of the Ni$^{2+}$ spin is then well defined along $z$ in both the ground and excited states involved in the optical transition. As a consequence, the amplitude of the replica transitions associated with a change in the Ni$^{2+}$ spin projection gradually decreases.

In this regime, the hole–Ni$^{2+}$ exchange interaction is maximal for $S_z = \pm 1$ and vanishes for $S_z = 0$. Three main emission lines are observed for the bright exciton in each circular polarization: a central line associated with $S_z = 0$ and two outer lines corresponding to $S_z = \pm 1$. The increase in the splitting of these three exciton lines between $B_z = 0$ and $B_z \approx 4$ T (see Fig.~\ref{Fig:MapBX}) reflects the progressive enhancement of the hole–Ni$^{2+}$ exchange interaction as the Ni$^{2+}$ spin aligns with the magnetic field.

At $B_z = 9$ T, the measured splitting between the $S_z = \pm 1$ lines, $\delta_{\pm1} \approx 0.34$ meV, is determined by both the hole–Ni$^{2+}$ and electron–Ni$^{2+}$ exchange interactions, $\delta_{\pm1}=3I_{hNi}-I_{eNi}$. Neglecting the electron–Ni$^{2+}$ exchange interaction, which was shown to be weak in charged Ni$^{2+}$-doped QDs \cite{PRB2025}, yields $I_{hNi} \approx 115~\mu$eV.

The intensity distribution among the three bright-exciton lines also evolves with magnetic field and, at fixed $B_z$, depends on the excitation power. The excitation-power dependence of the $\sigma^+$ lines at $B_z = 9$ T is shown in Fig.~\ref{Fig:PW}. At low excitation power, the PL is dominated by the low-energy line, associated with the most populated Ni$^{2+}$ spin state $S_z = -1$, corresponding to the state $\vert \Uparrow,\downarrow \rangle \vert -1 \rangle$. This behavior is consistent with the antiferromagnetic hole–Ni$^{2+}$ coupling previously identified in charged Ni$^{2+}$-doped QDs. The high-energy line, associated with $S_z = +1$, becomes visible only at higher excitation power.

This redistribution of intensities reflects an increase in the effective spin temperature, T$_{eff}$ of the magnetic ion at high excitation power. Such effects are commonly observed in dilute magnetic semiconductors \cite{Kneip2006}, particularly for magnetic ions with strong spin–strain coupling, which are sensitive to interactions with non-equilibrium phonons generated by high-energy optical excitation \cite{Tiwari2020}.

\section{Calculated magnetic field dependence of X-Ni$^{2+}$ and dark excitons recombination}

The PL spectra of X–Ni$^{2+}$ are calculated using the Hamiltonians $\mathcal{H}_{Ni}$ and $\mathcal{H}_{X,Ni}$. The optical transition probabilities are obtained from the squared matrix elements $|\langle S_z \vert X, S_z \rangle|^2$. The occupation of the X–Ni$^{2+}$ levels is described by the effective temperature $T_{\mathrm{eff}}$, which accounts for both the influence of the lattice temperature and the influence of optical excitation power via interaction with non-equilibrium phonons or spin-spin coupling.

\subsection{Calculated B$_z$ dependence of the PL of X-Ni$^{2+}$}

\begin{figure}[hbt]
\centering
\includegraphics[width=1.0\linewidth]{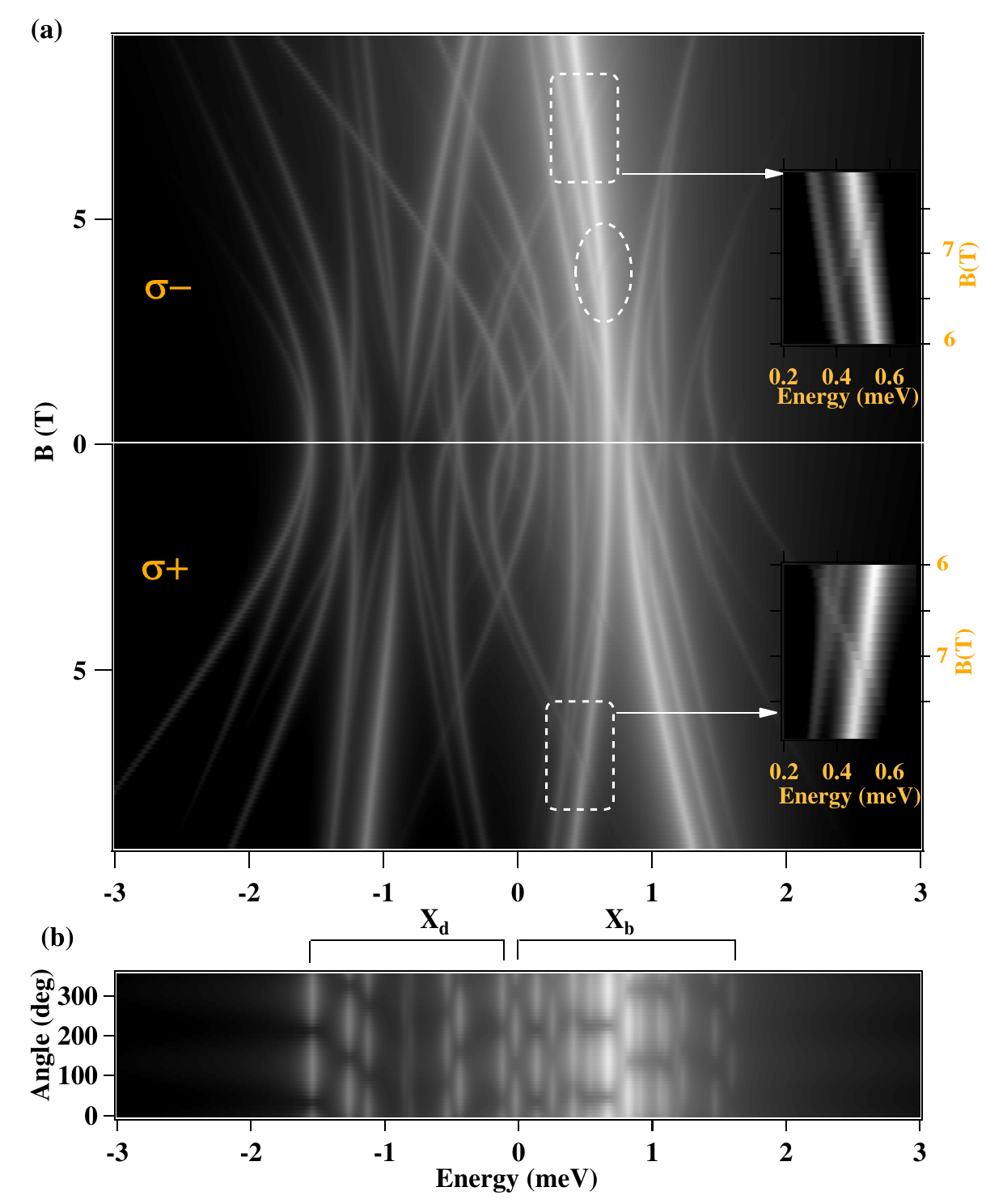}
\caption{(a) Calculated PL intensity map of the magnetic-field dependence of a neutral exciton in a Ni$^{2+}$-doped QD, obtained using the parameters listed in Table~\ref{TableParX}. The insets show details of dark-bright excitons anti-crossings. (b) Calculated linearly polarized PL intensity map at $B_z = 0$ T. The intensity maps are displayed on a logarithmic scale. The effective temperature is set to $T_{\mathrm{eff}} = 10$ K, and the emission lines are broadened with a Lorentzian profile of full width at half maximum (FWHM) of 50~$\mu$eV. Zero energy corresponds to an exciton with no exchange interaction.}
\label{Fig:modBX}
\end{figure}

Figure~\ref{Fig:modBX}(a) presents the calculated dependence of the X–Ni$^{2+}$ emission on $B_z$, using the parameters listed in Table~\ref{TableParX}. With this parameter set, the main features of the experimental spectra are reproduced. In particular, as it can be seen in Fig.~\ref{Fig:modBX}(b), at zero magnetic field the calculated PL is dominated by two groups of lines linearly polarized along orthogonal directions.

\begin{table}[!hbt]
    \center
\begin{tabular}{c|ccccccccccccc}
   Ni         & I$_{hNi}$ & I$_{eNi}$ & D$_{0}$ & E &  g$_{Ni}$ &  $\vartheta_{s}$ & $\phi_{s}$ & $\psi_{s}$ & $\eta$ \\
            & $\mu$eV   & $\mu$eV   & meV     & meV &  &  deg & deg & deg & \\
            \hline 
  & 100  & -50 & 0.5 & 0 & 2.0 &  45 & 0 & 0 & 0.4\\
            \hline \hline
   X           & $\frac{2}{3}\delta_{0}$  & $\Delta_{lh}$ & $\rho_{vb}$ & $\theta_{vb}$ & $\delta_1$ & $\gamma$  & g$_e$  & g$_h$  & $\delta_{xz}$ & $\delta_{yz}$ \\
            & meV   &        meV    &     meV  &  deg & $\mu$eV    & $\frac{ \mu eV}{T^{2}}$   &   & &   meV  &    meV   \\
            \hline 
         & -1.1 & 25 &   0.5 &  45 &  100  &   2   & -0.2   &  0.7 & 0 & 0 \\
            \hline 
\end{tabular}
    \caption{Parameters used in the modeling of the magnetic-field dependence of neutral-exciton PL spectra shown in Fig.~\ref{Fig:modBX}.}
    \label{TableParX}
\end{table}

Because of the large number of parameters involved, they cannot be adjusted independently, making a quantitative fit non-trivial. Nevertheless, $\delta_0$ can be estimated from the energy separation between the centers of the bright- and dark-exciton PL spectra. Values of $I_{hNi} = 100~\mu$eV and $I_{eNi} = -50~\mu$eV are chosen to approximate the total splitting of the three bright-exciton lines at high magnetic field $\delta_{\pm1}$.

For the bright exciton, both the long-range electron–hole exchange interaction and valence-band mixing induced by in-plane anisotropy contribute to the fine-structure splitting and linear polarization. Since these contributions cannot be determined independently, we use parameter values typical for CdTe/ZnTe QDs to reproduce the zero-field splitting and polarization properties of the bright exciton (see table \ref{TableParX}) \cite{Leger2007}.

In the calculations, the weaker satellite lines also appear on both sides of the main bright-exciton transitions. These lines correspond to recombination processes accompanied by a change in the Ni$^{2+}$ spin state. Such replicas emerge in the calculated spectra when a rotation of the strain tensor, characterized by the angle $\vartheta_s$, is included. For simplicity, we set $\varphi_s = \psi_s = 0$. A value of $D_0 = 0.5$ meV is chosen, as it approximately reproduces the experimental positions of the satellite lines for $\vartheta_s = 45^\circ$. These parameters are consistent with those reported for charged Ni$^{2+}$-doped QDs \cite{PRB2025}. As shown below, they are also compatible with the observed splitting of the dark exciton states. In the PL spectra, the dominant radiative channel of the bright exciton, which conserves the Ni$^{2+}$ spin, limits the contribution of these weakly allowed replica transitions.

With increasing $B_z$, the model also reproduces the progressive transformation of the two linearly polarized bright-exciton lines at zero field into three lines in each circular polarization. The intensity distribution among these three lines evolves with magnetic field. At high field, the intensity maximum occurs on the low-energy line in $\sigma^+$ polarization and on the high-energy line in $\sigma^-$ polarization. This behavior results from thermalization of the Ni$^{2+}$ spin into the $S_z = -1$ state, as described by the effective spin temperature $T_{\mathrm{eff}}$, combined with the antiferromagnetic hole–Ni$^{2+}$ exchange interaction.

The calculated spectra also show that the relative intensity of the satellite lines on both sides of the bright-exciton transitions decreases with increasing $B_z$. This reduction is expected as a strong longitudinal magnetic field restores a common quantization axis for the initial and final states, thereby suppressing transitions involving changes in the Ni$^{2+}$ spin state.

\subsection{PL of dark excitons in a Ni$^{2+}$-doped QD}

The model provides more insight into the optical transitions observed on the low-energy side of the experimental PL spectra. Under a longitudinal magnetic field $B_z$ (Fig.~\ref{Fig:MapBX}), these transitions evolve into a broad fan-like structure. A quantitative fit of the main transition energies is shown in Fig.~\ref{Fig:PW}(a). Such weak-intensity lines appear in the calculated spectra and originate from recombination of the dark excitons, which are shifted from the bright excitons by the electron–hole exchange interaction. The characteristic fan-like pattern is well reproduced in PL intensity map calculated with the parameters of Table \ref{TableParX} (Fig.~\ref{Fig:modBX}(a)), supporting the consistency of the model.

To first approximation, all X$_d$ optical transitions are forbidden. Unlike the bright exciton, there is no fast recombination channel that conserves the Ni$^{2+}$ spin (as in the high-intensity, linearly polarized central lines of X$_b$). Optical recombination therefore requires mixing with a bright-exciton state via either a hole spin flip or an electron spin flip.

In the model, several mechanisms can induce bright–dark exciton mixing. A first contribution arises from carrier–Ni$^{2+}$ flip-flop processes associated with their exchange interaction. Such flip-flops are naturally present in the electron–Ni$^{2+}$ coupling, $I_{eNi}\vec{\sigma}\cdot\vec{S}$. In contrast, hole–Ni$^{2+}$ flip-flops occur only when valence-band mixing induced by in-plane strain anisotropy is included (Appendix A). Additionally, valence-band mixing caused by shear strain, combined with the short-range electron–hole exchange interaction, enables electron spin flips that conserve the hole spin. This mechanism mixes the $+2$ and $+1$ exciton states on one side, and the $-2$ and $-1$ states on the other. Finally, a hole spin flip conserving the Ni$^{2+}$ spin is introduced by the additional $\eta$ term in the exchange interaction. This term couples the bright excitons $\pm 1$ to the dark excitons $\mp 2$ respectively and conserves the Ni$^{2+}$ spin S$_z$.

\subsubsection{Identification of the dark exciton lines}

To elucidate the origin of the various optical transitions of the dark-exciton, we first note that the overall splitting of its PL spectrum is significantly larger than that of the bright exciton. As in the bright-exciton case, the outer lines of the dark-exciton spectrum arise from recombination processes involving a change in the Ni$^{2+}$ spin level. In contrast to the bright exciton, however, these replica transitions dominate the dark-exciton emission or are at least comparable in intensity to the central lines. This behavior is particularly pronounced at zero or weak B$_z$, where the central spin-conserving transitions are suppressed. As presented in Fig.~\ref{Fig:modBX}(b), within the framework of the model, six main optical transitions are expected for the dark exciton at zero magnetic field.

\begin{figure}[hbt]
\centering
\includegraphics[width=1\linewidth]{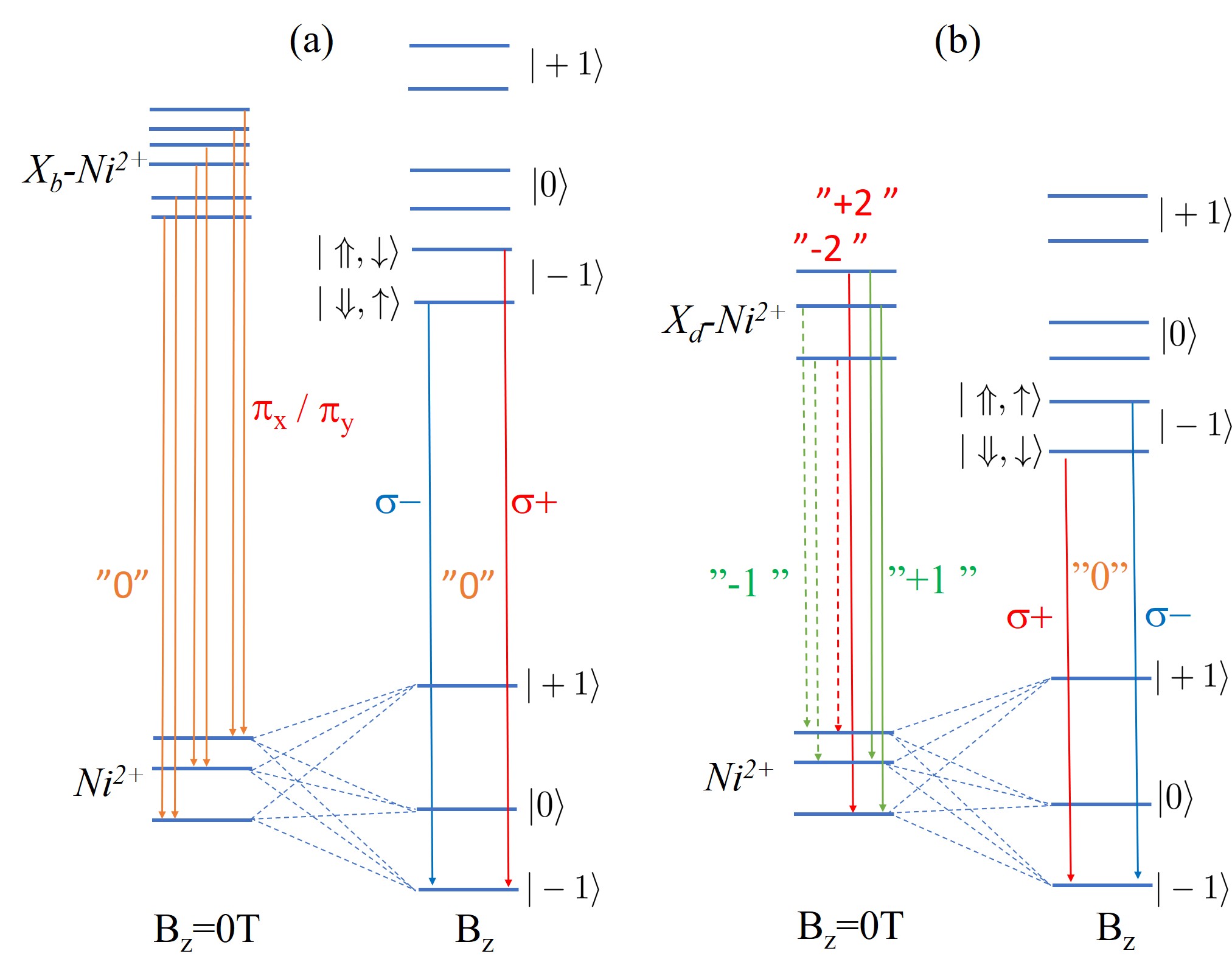}
\caption{Energy-level scheme at zero magnetic field and under a large $B_z$ for the initial (X–Ni$^{2+}$) and final (Ni$^{2+}$) states of a neutral Ni$^{2+}$-doped QD in the presence of strain misorientation. The dominant recombination pathways for the bright and dark excitons are shown in panels (a) and (b), respectively. "i" denotes an optical transition involving a change in the Ni$^{2+}$ level of i.}
\label{Fig:scheme}
\end{figure}

The replicas can involve changes of up to two units in the Ni$^{2+}$ spin projection (see the energy-level scheme in Fig.~\ref{Fig:scheme}). Two transitions correspond to double spin-flip processes, one on the low-energy side and one on the high-energy side and give rise to the outermost lines of the dark-exciton spectrum  (Fig.~\ref{Fig:scheme}(b)). Their total zero-field splitting is governed by the strain-induced fine-structure parameters $D_0$ and $E$, as well as by the rotation angles of the local strain frame, primarily $\vartheta_s$. Such transitions are not observed in the bright-exciton spectra, which explains the smaller overall splitting for the bright excitons compared to the dark excitons.

The lines closer to the center of the dark-exciton spectrum correspond either to single spin-flip processes or to spin-conserving transitions. Four optical transitions involve a change of one unit in the Ni$^{2+}$ spin projection, with two located at higher energy and two at lower energy relative to the spin-conserving transitions (see Fig.~\ref{Fig:scheme}(b)). The splitting of both the high- and low-energy replicas into two distinct lines originates from the non-equidistant splitting of the three Ni$^{2+}$ spin levels in the initial (X–Ni$^{2+}$) and final (Ni$^{2+}$) states involved in the optical transition.

\subsubsection{Evolution of dark excitons at low B$_z$}

Magnetic fields below 2~T already induce a redistribution of oscillator strengths among the dark-exciton transitions. In particular, the spin-conserving recombination lines located at the center of the PL structure of the dark exciton are nearly suppressed at $B_z=0$, while their intensity increases for $B_z\approx$1–2~T. This field-induced recovery is clearly visible in the calculated spectra of Fig.~\ref{Fig:modBX}. 

As observed for the bright excitons, under B$_z$, the PL is dominated by states associated with S$_z$=-1 which is thermally favored. Within the model, the dark state $\vert\Downarrow,\downarrow\rangle\vert -1\rangle$ (central line shifting towards low energy under B$_z>$0) acquires oscillator strength in $\sigma^+$ polarization, while the state $\vert\Uparrow,\uparrow\rangle\vert -1\rangle$ (central line shifting towards high energy under B$_z>$0) becomes visible in $\sigma^-$ polarization. Their finite oscillator strength results from mixing with the bright excitons $\vert\Uparrow,\downarrow\rangle$ and $\vert\Downarrow,\uparrow\rangle$, respectively. This bright–dark coupling is mediated by the $\eta$ term in the exchange interaction (spin flip of the hole conserving the Ni$^{2+}$ spin). The corresponding dark-exciton lines exhibit a magnetic-field dispersion opposite to that of the main bright-exciton branches, with the $\sigma^-$ line shifting toward higher energy and the $\sigma^+$ line toward lower energy.

The model further shows that the minimum intensity of the Ni$^{2+}$ spin-conserving transitions does not occur at $B_z=0$, but is shifted to approximately 1~T in $\sigma^+$ polarization. In this low-field regime, the Zeeman splitting of the Ni$^{2+}$ ion, determined by its $g$ factor, governs the evolution of both excited- and ground-state energies.

The bright--dark coupling term proportional to $\eta$ depends explicitly on $S_z$. For spin-conserving transitions, the initial and final states possess identical $S_z$ compositions. When the strain-induced mixing is maximal, $\langle S_z\rangle \approx 0$, which leads to a cancellation of the bright components in the overlap between the initial and final states and therefore to a vanishing oscillator strength. 

Maximal strain-induced mixing occurs at the magnetic field for which the external Zeeman field compensates the hole--Ni$^{2+}$ exchange field acting on the magnetic ion. The value of this compensation field is determined by the exchange integral $I_{h\mathrm{Ni}}$. As $B_z$ increases further, the strain-induced mixing is progressively suppressed and the eigenstates approach pure $S_z$ projections. Consequently, $\langle S_z\rangle$ reaches its maximum value and the recombination channel controlled by the $\eta$-induced bright--dark mixing becomes effective.

By contrast, for the outer dark-exciton lines, which involve a change in the Ni$^{2+}$ spin levels, the field dependence of the oscillator strength is qualitatively different. These transitions are strong at zero or weak magnetic fields and progressively weaken as $B_z$ increases. 

Their finite oscillator strength originates from the overlap between initial and final states characterized by different strain-induced $S_z$ mixing. Because the mixing coefficients in these states differ, the cancellation of the bright components does not occur, and a finite oscillator strength is therefore maintained at zero or weak magnetic fields.

A weak longitudinal magnetic field also slightly affects the overall splitting of the dark excitons. Experimentally, the splitting is $\delta_d \approx 1.55$ meV at $B_z = 0$ and reaches a minimum value of $\delta_d = 1.47$ meV at $B_z \approx 0.5$ T in $\sigma^+$ polarization. This reduction of the splitting is also clearly reproduced by the model.

The reduction of the spectral width under weak $B_z$ originates from a modification of the hole–Ni$^{2+}$ exchange interaction. When the total effective field (external plus exchange field) vanishes, the mixing of the $S_z$ states induced by $\mathcal{H}_{\mathrm{Ni}}$ is maximal. As a result, the average value of $S_z$ in each eigenstate is close to zero, leading to a minimal energy shift of the exciton due to the exchange interaction with the magnetic atom. In the ground state, the Ni$^{2+}$ levels are shifted by the Zeeman energy associated with the corresponding external magnetic field. Reducing the hole–Ni$^{2+}$ exchange interaction lowers the energy of the high-energy dark-exciton level and consequently shifts the corresponding optical transition to lower energy. This is the origin of the curvature observed in the high-energy line's low-magnetic-field dependence, which is clearly seen in the experiment and reproduced by the model.

\subsubsection{Dark excitons under large B$_z$, anticrossings}

At large $B_z$, the Zeeman splitting of the Ni$^{2+}$ spin progressively overcomes the strain-induced mixing described by $\mathcal{H}_{\mathrm{Ni}}$, and the spin levels approach pure $S_z$ eigenstates. Experimentally, this alignment is reflected in the bright-exciton spectra by the increasing separation of the three emission lines in each circular polarization.

Although dark–bright mixing mediated by the coupling constant $\eta$ persists at high fields, it conserves $S_z$. As the Ni$^{2+}$ eigenstates become nearly pure, the overlap between different spin levels in the initial and final states vanishes, leading to a suppression of the replica lines. The dark exciton can nevertheless recombine via the central transition without changing the Ni$^{2+}$ spin level, owing to the hole spin flip term $\eta$.

The Zeeman splittings of selected dark-exciton lines significantly exceed those of a bare exciton, as they incorporate the Zeeman contribution of the magnetic ion. In particular, the lowest- and highest-energy dark-exciton branches exhibit effective $g$ factors larger than that of the bright exciton, consistent with transitions involving spin flips of the Ni$^{2+}$ ion.

Around $B_z \approx 7.5$~T, an anticrossing is observed between the high-energy bright exciton in $\sigma^-$ polarization and a dark-exciton branch. A corresponding anticrossing is also visible in $\sigma^+$ polarization, involving the highest-energy dark-exciton level at the same magnetic field.

This feature originates from the coupling between the bright-exciton state $\vert\Downarrow,\uparrow\rangle\vert -1\rangle$ and a dark-exciton state that becomes resonant at this field. It appears in the model that the magnitude of the anticrossing gap is controlled by the electron–Ni$^{2+}$ exchange interaction $I_{e\mathrm{Ni}}$. In the calculated spectra, a gap starts to be observed for $I_{e\mathrm{Ni}}$ values on the order of a few tens of $\mu$eV. The coupling mechanism is then an electron–Ni$^{2+}$ flip-flop process that mixes bright and dark exciton configurations. It involves an electron spin flip while conserving the hole spin, coupling the bright state $\vert\Downarrow,\uparrow\rangle\vert -1\rangle$ to the component $\vert\Downarrow,\downarrow\rangle\vert 0\rangle$ of the overlapping dark-exciton. At lower magnetic fields, the associated dark-exciton branch connects continuously to transitions involving a change of two units in the Ni$^{2+}$ spin projection. The anticrossing observed around 3 T on the high-energy line in $\sigma^-$ polarization (see Fig.~\ref{Fig:MapBX}) is also reproduced by the model. We emphasize that the energy gap at the anticrossings does not provide a direct measurement of the exchange integral $I_{e\mathrm{Ni}}$, as it also depends on the S$_z$ mixing of the interacting levels. Nevertheless, the occurrence of anticrossings at magnetic fields consistent with the experimental values further supports the validity of the model.

\section{Conclusion}

We have investigated the magneto-optical properties of a neutral exciton coupled to the spin of a single Ni$^{2+}$ ion in a CdTe/ZnTe QD. The PL spectra reveal that the orientation of the local strain tensor at the position of the magnetic atom plays a central role in determining the optical signatures of the exciton–spin interaction. A misalignment between the strain principal axis and the QD growth direction mixes the Ni$^{2+}$ spin states and strongly reduces the effective hole–Ni$^{2+}$ exchange interaction at low magnetic field. This strain-induced spin mixing gives rise to characteristic satellite lines around the bright-exciton transitions and strongly influences the recombination of dark excitons.

The application of a longitudinal magnetic field progressively restores a well-defined spin quantization axis and the circularly polarized optical selection rules, allowing the three Ni$^{2+}$ spin projections to be resolved in the emission spectra. The dark-exciton emission is dominated at low magnetic field by transitions involving spin flips of the magnetic ion, producing a characteristic fan-like pattern whose magnetic-field evolution involves the Ni$^{2+}$ g-factor.

Our results demonstrate that the local strain environment can control the effective spin eigenstates of transition-metal dopants in semiconductor QDs. This sensitivity of Ni$^{2+}$ spins to strain makes such systems promising platforms for hybrid spin–mechanical quantum devices and for strain-based control of individual magnetic ions.


\appendix

\section{Valence-band mixing in a strained quantum dot}

The lh–hh mixing induced by shape or strain anisotropy is described by the valence-band Hamiltonian $\mathcal{H}_{VB}$ written in the basis ($|\frac{3}{2},+\frac{3}{2}\rangle,|\frac{3}{2},+\frac{1}{2}\rangle,|\frac{3}{2},-\frac{1}{2}\rangle,|\frac{3}{2},-\frac{3}{2}\rangle$) \cite{Chuang,Voon2009}:

\begin{equation}
\mathcal{H}_{VB} =
\begin{pmatrix}
0 & Q & P & 0\\
Q^* & \Delta_{lh} & 0 & P\\
P^* & 0 & \Delta_{lh} & -Q\\
0 & P^* & -Q^* & 0
\end{pmatrix}.
\label{HBM}
\end{equation}

Here $P=\delta_{xx,yy}+i\delta_{xy}$ describes in-plane anisotropy and $Q=\delta_{xz}+i\delta_{yz}$ a symmetry breaking in the $xz$ or $yz$ plane. These terms can originate from shape anisotropy (Kohn–Luttinger Hamiltonian) or strain (Bir–Pikus Hamiltonian) \cite{Voon2009}. The in-plane mixing parameter is written $P=\rho_{vb}e^{-2i\theta_{vb}}$, where $\rho_{vb}$ is the mixing amplitude and $\theta_{vb}$ the principal anisotropy axis measured from [100] \cite{PRB2025,Tiwari2021}.

With in-plane anisotropy, $P\neq0$ and for weak mixing ($\rho_{vb}\ll\Delta_{lh}$), the hole ground states read

\begin{eqnarray}
|\Phi_h^+\rangle &=& |+3/2\rangle-\phi|-1/2\rangle, \nonumber\\
|\Phi_h^-\rangle &=& |-3/2\rangle-\phi^*|+1/2\rangle ,
\end{eqnarray}

with $\phi=(\rho_{vb}/\Delta_{lh})e^{2i\theta_{vb}}$.  
A first order development of the angular momentum operator on the $\{|\Phi_h^\pm\rangle\}$ subspace gives

\begin{eqnarray}
\tilde{j}_+ &=&
\frac{\rho_{vb}}{\Delta_{lh}}
\begin{pmatrix}
0 & -2\sqrt{3}e^{-2i\theta_{vb}}\\
0 & 0
\end{pmatrix}, \nonumber\\
\tilde{j}_- &=&
\frac{\rho_{vb}}{\Delta_{lh}}
\begin{pmatrix}
0 & 0\\
-2\sqrt{3}e^{2i\theta_{vb}} & 0
\end{pmatrix}, \\
\tilde{j}_z &=&
\begin{pmatrix}
3/2 & 0\\
0 & -3/2
\end{pmatrix}. \nonumber
\end{eqnarray}

These terms enable heavy-hole spin flips. Combined with the short-range electron–hole exchange interaction $(2/3)\delta_0(\vec{\sigma}\!\cdot\!\vec{J})$, they induce mixing of the bright excitons and allow electron–hole flip-flop processes. In magnetic QDs, similar hh–Ni$^{2+}$ flip-flops can occur.

For a distortion in a vertical plane ($Q\neq0$), the hole states become

\begin{eqnarray}
|\Xi_h^+\rangle &=& |+3/2\rangle+\xi|+1/2\rangle, \nonumber\\
|\Xi_h^-\rangle &=& |-3/2\rangle-\xi^*|-1/2\rangle .
\end{eqnarray}

A first order development of the angular momentum operator on the $\{|\Xi_h^\pm\rangle\}$ subspace yields

\begin{eqnarray}
\tilde{j}_+ &=& 
\xi
\begin{pmatrix}
\sqrt{3} & 0\\
0 & -\sqrt{3}
\end{pmatrix}, \nonumber\\
\tilde{j}_- &=&
\xi^*
\begin{pmatrix}
\sqrt{3} & 0\\
0 & -\sqrt{3}
\end{pmatrix}, \\
\tilde{j}_z &=&
\begin{pmatrix}
3/2 & 0\\
0 & -3/2
\end{pmatrix}. \nonumber
\end{eqnarray}

This mixing allows the hole–Ni$^{2+}$ exchange interaction to couple $|\Xi_h^{\pm},S_z\rangle$ to $|\Xi_h^{\pm},S_z\pm1\rangle$, corresponding to a Ni$^{2+}$ spin flip while conserving the hh spin projection. Combined with the electron–hole exchange interaction, it further couples the $|+1\rangle$ and $|+2\rangle$ excitons (and similarly $|-1\rangle$ and $|-2\rangle$) \cite{Tiwari2021}.

\end{document}